\documentclass[sigconf]{acmart}

\usepackage{subcaption}
\usepackage{multirow}
\usepackage{makecell, enumitem}
\usepackage{colortbl}

\definecolor{markerYellow}{RGB}{235,190,20}
\definecolor{markerBlue}{RGB}{40,105,210}
\definecolor{markerRed}{RGB}{205,50,50}
\definecolor{markerMagenta}{RGB}{190,45,180}

\copyrightyear{2026}
\acmYear{2026}
\setcopyright{cc}
\setcctype{by}
\acmConference[SUI '26]{ACM Symposium on Spatial User Interaction}{October 10--11, 2026}{Bari, Italy}
\acmBooktitle{ACM Symposium on Spatial User Interaction (SUI '26), October 10--11, 2026, Bari, Italy}
\acmDOI{10.1145/3822518.3830045}
\acmISBN{979-8-4007-2812-9/2026/10}

\begin{document}

\title[Code Black]{Code Black: Desktop-Mediated Co-Design of AR-HMD Microinteractions for Emergency Department Teamwork}

\author{Jonathan Segal}
\orcid{0000-0002-8506-3785}
\affiliation{%
  \institution{Cornell University}
  \city{New York}
  \state{NY}
  \country{USA}
  \postcode{10044}}
\email{jis62@cornell.edu}

\author{Jalynn Nicoly}
\orcid{0000-0002-2897-5833}
\affiliation{%
  \institution{Colorado State University}
  \city{Fort Collins}
  \state{CO}
  \country{USA}
  \postcode{80523}}
\email{Jalynn.Nicoly@colostate.edu}

\author{Francisco Ortega}
\orcid{0000-0002-2449-3802}
\affiliation{%
  \institution{Colorado State University}
  \city{Fort Collins}
  \state{CO}
  \country{USA}
  \postcode{80523}}
\email{fortega@colostate.edu}

\author{Angelique Taylor}
\orcid{0000-0003-1285-6431}
\affiliation{%
  \institution{Cornell University}
  \city{New York}
  \state{NY}
  \country{USA}
  \postcode{10044}}
\email{amt298@cornell.edu}

\renewcommand{\shortauthors}{Segal et al.}
\settopmatter{authorsperrow=4}

\begin{abstract}
Emergency Department (ED) teams coordinate shifting roles, medication decisions, and time-critical interventions under uncertainty.
Augmented reality head-mounted displays (AR-HMDs) have shown potential to spatially anchor information during care, creating opportunities to examine how spatial interfaces might support teamwork. We conducted a speculative co-design study with 12 healthcare workers (HCWs) using an editable, desktop-mediated Unity-based 3D design probe to visualize and refine \textit{work-as-imagined} AR-HMD interfaces for role-based notifications, task-specific timers, and dosage verification. 
Guided by microinteraction rules, participants identified future spatial user interfaces (SUI) requirements such as how they appear, update, or are dismissed in relation to clinical practice, safety concerns, and existing tools.
Five returning participants and 26 additional HCWs subsequently provided follow-up feedback on derived visual interface alternatives.
Findings show that desktop-mediated spatial co-design elicited formative specifications for role visibility, task-linked timing, and verification-oriented dosage assistance, while revealing tensions involving clutter, shared awareness, communication, privacy, and reliability.
Rather than evaluating a functional AR-HMD system or team-based clinical performance, this study contributes the Speculative Co-Design Framework for AR-HMD Teamwork (SCF-HMD) and a visual design catalog for translating expert critique of work-as-imagined (WAI) concepts into situated goals for future AR-HMD systems.
\end{abstract}

\begin{teaserfigure}
\centering
\includegraphics[width=0.85\textwidth]{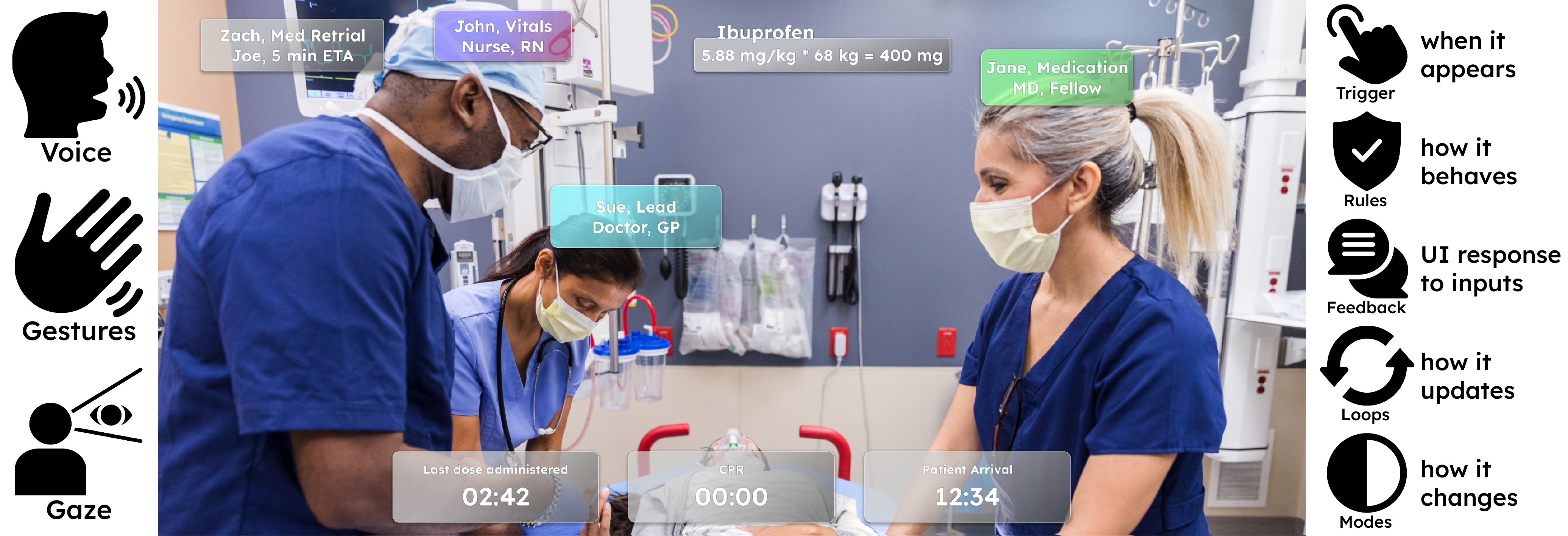}
\caption{XR visual showing role-based notifications, task-specific timers, and dosage verification interfaces. Participants reviewed an image, not a functional AR-HMD system. {\scriptsize Photo: \href{https://www.istockphoto.com/photo/medical-staff-works-together-to-obtain-vitals-gm1421626432-467193863}{iStock/SDI Productions}.}}
\Description{Composite visual interface alternative showing clinicians in an emergency care environment with overlaid role-based notifications, task-specific timers, and dosage verification interfaces.}
\label{fig:teaser}
\end{teaserfigure}

\ccsdesc[500]{Human-centered computing~User interface design}
\ccsdesc[500]{Human-centered computing~Participatory design}
\ccsdesc[300]{Human-centered computing~Interface design prototyping}
\ccsdesc[300]{Human-centered computing~User centered design}

\keywords{spatial user interfaces, augmented reality, head-mounted displays, participatory design, emergency medicine, healthcare teamwork}

\maketitle


\section{Introduction} 
\label{Introduction}

In the Emergency Department (ED), healthcare workers (HCWs) rapidly coordinate in teams to treat critically ill patients in time-sensitive, safety-critical environments \cite{kilner_role_2010, kusunoki_sketching_2015}.
Effective coordination depends on maintaining a common ground of the patient’s status, team members’ responsibilities, and the actionable interventions as the care team evolves \cite{sarcevic_coordinating_2011}.
However, clinical teams struggle to maintain situational awareness and experience high workloads due to breakdowns in communication, coordination, and information access \cite{warden_information_2024, warden_information_2022, blackburn_information_2019, pun_factors_2015}.
To address this challenge, ED teams rely on established practices including Team Strategies and Tools to Enhance Performance and Patient Safety (TeamSTEPPS) \cite{cooke_teamstepps_2016} and ambient displays \cite{khazaal_highly_2022}, to make responsibilities and patient information visible during care \cite{mastrianni_beyond_2025, zellner_simple_2024}.
Augmented reality head-mounted displays (AR-HMDs) have shown potential to spatially anchor information to people, tasks, and equipment creating opportunities to examine how spatial interfaces might support teamwork. Although AR-HMDs have been explored for surgery \cite{desselle_augmented_2020, cofano_augmented_2021}, clinical operations \cite{klinker_digital_2020}, and training \cite{gerup_augmented_2020, parsons_current_2021}, limited research has examined the design of spatial user interfaces (SUIs) for ED teamwork.
This presents a new design problem focused on how spatial interfaces present information without obstructing care, disrupting shared situational awareness, or duplicating existing tools.
Therefore, we examine how AR-HMD interfaces can complement existing ED tools and practices for teamwork using design methodologies.

AR-HMD interfaces for HCW teams must account for patient state, human actions, environmental conditions, shared awareness, and situated clinician control.
Researchers have largely examined AR-HMDs in healthcare settings where one user completes a defined clinical task \cite{mamone_monitoring_2020, kuge_design_2021}, with emerging work examining collaborative AR-HMD use in settings such as surgical teleconsulting \cite{maria_supporting_2023}. 
Yet, as ED care teams dynamically assemble, shift roles and responsibilities, information access needs change and evidence-based practices shape actionable interventions. AR-HMD design must account for not only single-user interactions but also multi-user collaboration \cite{blackburn_information_2019, kusunoki_designing_2015, sarcevic_coordinating_2011, schafer_survey_2022, radu_survey_2021}.
However, open questions remain about how clinicians can inform the design of spatial interface behavior for team interactions in safety-critical environments.

Design methods such as participatory design \cite{kristensen_participatory_2006}, Research-through-Design (RtD) \cite{zimmerman_research_2007}, and speculative design \cite{blythe_research_2014} provide opportunities to investigate design futures, including pre-deployment critiques of future SUIs.
In emergency care, participatory design has involved domain experts in envisioning technologies for future emergency medical practice \cite{kristensen_participatory_2006, taylor_hospitals_2022}.
Extended Reality (XR) research, in particular, has employed immersive design fictions to situate speculative interfaces within virtual environments \cite{mcveigh-schultz_immersive_2018}, participatory design fiction with mixed reality to elicit feedback on future interfaces \cite{chi_participatory_2022}, and iterative participatory design to develop collaborative AR activities \cite{ullal_iterative_2024}.
By making future interfaces visible, these methods allow stakeholders to critique the alignment of proposed designs with work as done (WAD) before functional development.
These approaches are particularly relevant to the distinction between \textit{work-as-imagined} (WAI)--envisioned future systems and procedures, and \textit{work-as-done} (WAD)--real-world constraints \cite{braithwaite_resilient_2017}.
This distinction motivates a design framework using desktop-mediated 3D design probes and microinteraction prompts to bridge WAI and WAD, helping researchers iteratively refine speculative SUI concepts before implementation and deployment.

Although speculative and participatory XR design probes enable pre-deployment critique of future interfaces \cite{colombo_augmented_2018, mcveigh-schultz_immersive_2018, chi_participatory_2022}, several research gaps remain.
First, limited work has examined how researchers can co-design AR-HMD interfaces for teams working under dynamic, safety-critical conditions such as the ED.
This gap matters because interface designs that appear useful in the absence of real-world conditions may conflict with clinicians' accounts of WAD, uncovering obstructive, redundant, or disruptive design concepts.
Second, prior XR research has examined moment-to-moment interface behaviors such as spatial notifications, collaborative attention, and interaction modalities \cite{plabst_exploring_2023, piumsomboon_effects_2019}, but provides limited guidance on how to elicit these behaviors during co-design.
This is important because HCWs must reason about AR-HMD interaction including how it becomes visible, remains useful, changes over time, and supports clinician control during care.
Third, while understanding the needs of HCWs is important, accomplishing this goal alone does not enable researchers to build SUIs; thus, there is a need for methodological frameworks that characterize technological interventions that bridge the WAI and WAD gap.

To address these challenges, we conducted a speculative co-design study with 12 HCWs using an editable desktop-mediated Unity-based 3D design probe of future AR-HMD interfaces for ED teamwork to reflect on spatial interface elements in relation to clinicians, equipment, and existing coordination practices.
Using microinteractions as a framework, we elicited design requirements for future AR-HMD interfaces that respond to the practical demands of team-based emergency care.
Five returning participants and 26 additional HCWs later provided follow-up feedback on visual interface alternatives derived from the co-design sessions.
Rather than evaluating a deployed AR-HMD system, our study investigates how a spatial co-design approach can elicit formative interaction specifications for future team-oriented SUIs in safety-critical care.

Our \textbf{contributions} are:
\begin{enumerate}
\item A \textbf{formative desktop-mediated spatial co-design method for bridging the WAI and WAD gap for AR-HMD interfaces} that employs a Unity-based clinical scene with speculative prompts to ground HCWs in ED spatial interfaces based on existing practice, constraints, and concerns.
\item The \textbf{Speculative Co-Design Framework for AR-HMD Teamwork (SCF-HMD)}, which supports the elicitation of situated SUI specifications by combining WAI/WAD framing, iterative refinement of low-fidelity spatial concepts, and a microinteraction framework.
\item A \textbf{visual design catalog of speculative AR-HMD interface concepts for ED teamwork} that synthesizes the design concepts refined during the co-design sessions.\footnote{\url{https://heyzine.com/flip-book/5206d9f05b.htm}} This catalog provides an artifact for examining how clinical experts envision spatial interfaces before implementation.
\end{enumerate}

\section{Background}
\label{Background}

\subsection{Teamwork in Healthcare}
Emergency care depends on teams coordinating under time pressure and uncertainty.
In trauma resuscitation and other ED procedures, team leaders manage shifting responsibilities among HCWs who may not regularly work together \cite{chellam_singh_operating_2023, kusunoki_designing_2015}.
These teams coordinate effectively when members share an understanding of the patient’s condition and the team’s next actions \cite{sarcevic_coordinating_2011}.
Closed-loop communication supports this shared understanding by prompting clinicians to assign, repeat, and confirm information \cite{salik_closed_2024}.

Untimely or inaccessible information can disrupt communication, delay action, and compromise patient safety \cite{pun_factors_2015, savioli_emergency_2022}.
Recent work on action teams reinforces this concern by showing that technologies for time-sensitive teamwork must provide information without adding distraction, workload, or communication burden \cite{tanjim_help_2025, taylor_rapidly_2025}.
Together, this work shows that ED team coordination depends on timely access to relevant information, but offers less guidance on when spatial, wearer-specific interfaces may complement rather than disrupt existing coordination practices.
This distinction is important because clinicians' accounts of WAD reflect time pressure, physical demands, interruptions, and constraints that speculative interface concepts cannot fully reproduce.

\subsection{AR-HMDs in Healthcare}
Researchers have explored AR-HMDs for presenting clinical information within the wearer's field of view \cite{mamone_monitoring_2020, kuge_design_2021, lawson_effect_2023, mastrianni_transitioning_2023}.
For example, OpticARe mirrors patient-monitor data through an AR-HMD, but offers less guidance for coordinating several clinicians around changing information \cite{kimmel_opticare_2021}.
Emerging work has examined collaborative AR-HMD use, including surgical teleconsulting \cite{maria_supporting_2023}, while broader collaborative Mixed Reality (MR) research shows how spatial cues communicate collaborators' attention and actions \cite{piumsomboon_effects_2019}.
However, high-intensity emergency care remains underexplored for designing spatial interfaces that provide coordination-relevant information without obscuring care, fragmenting shared awareness, or duplicating existing tools.
This gap motivates early-stage design approaches that let HCWs critique AR-HMD support before development or clinical evaluation.

\begin{figure}[t]
\centering
\includegraphics[width=\linewidth]{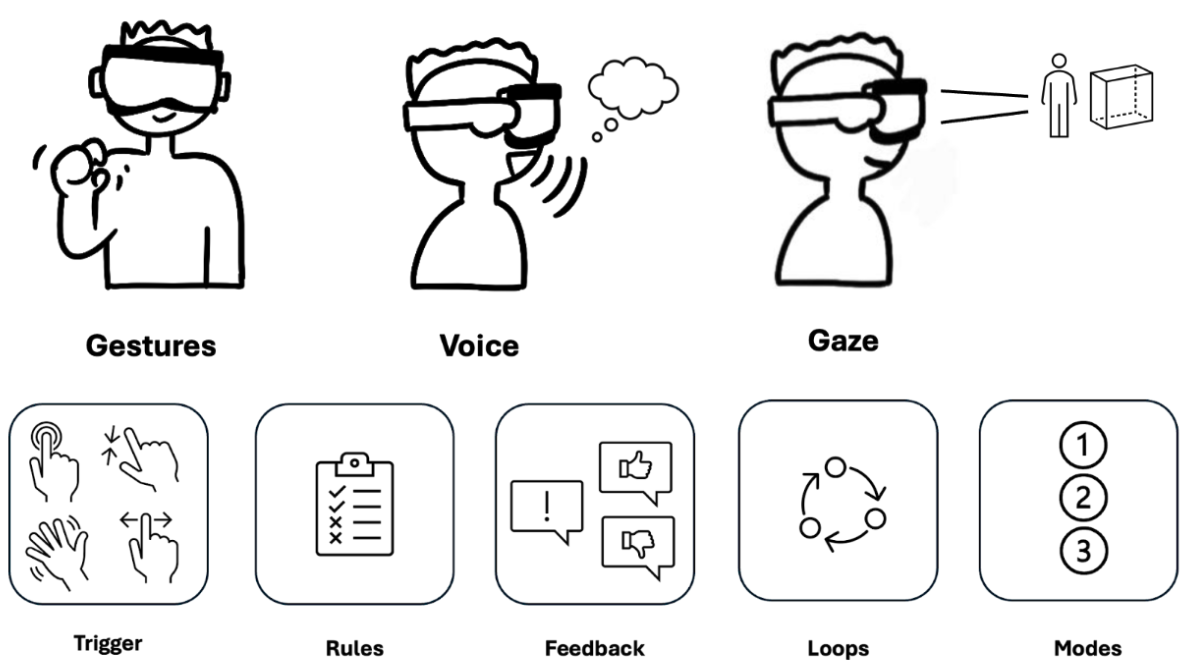}
\caption{Interaction modalities and microinteraction concepts introduced to participants during the interviews.}
\Description{The top row of the image illustrates interaction or trigger modalities for AR-HMDs, showcasing three primary methods: Gestures, where users interact via hand movements; Voice, where voice commands are used to control the system; and Gaze, where users engage with the environment through eye-tracking and focusing on specific objects. The bottom row represents microinteractions that follow these triggers, including Trigger icons for specific actions like tapping or swiping, Rules for governing system responses, Feedback through visual or auditory signals, Loops for continuous system updates, and Modes for managing different interaction settings.}
\label{fig:modalities}
\end{figure}

\subsection{Participatory Design in HCI}
\label{sec:pd_in_hci}

Participatory design supports healthcare technology design by positioning HCWs as domain experts who can assess whether prototypes align with clinical workflows, priorities, and safety constraints \cite{gheidar_integrating_2023, taylor_towards_2024}.
Within participatory design, co-design involves stakeholders in generating, critiquing, and refining technologies or design futures \cite{sanders_co-creation_2008}.
RtD, in contrast, produces knowledge through creating and reflecting on artifacts \cite{zimmerman_research_2007, luria_research_2021}.
Speculative design and design fiction represent technology futures that stakeholders can critique \cite{blythe_research_2014, lyckvi_role_2018}.
Design probes are artifacts that elicit reflection on proposed technologies and practices, including paper prototypes, digital 3D mockups, and virtual environments \cite{mattelmaki_design_2006, taylor_hospitals_2022, mcveigh-schultz_immersive_2018}.

Immersive design probes situate speculative interfaces in virtual reality (VR) and MR to elicit feedback on design futures \cite{mcveigh-schultz_immersive_2018, simeone_immersive_2022, chi_participatory_2022}.
Speculative MR prototypes have explored future collaborative technologies in synchronous groups, showing how spatial representations can elicit team-level reflection on emerging systems \cite{johnson_exploring_2025}.
In healthcare, prior work has envisioned emergency medical practices through participatory design \cite{kristensen_participatory_2006}, co-designed AR concepts with nurses \cite{albrecht-gansohr_playful_2023}, and applied RtD with design fiction and 3D mockups as probes for resilient EDs \cite{taylor_hospitals_2022}.
Together, this literature demonstrates the value of participatory, speculative, and spatial approaches for discussing future healthcare and collaborative technologies.
However, it offers limited guidance for co-designing AR-HMD interfaces for dynamic ED teamwork at the level of spatial placement and moment-to-moment behavior.

\subsection{Microinteractions in Spatial Interfaces}
\label{sec:microinteractions}
Microinteractions are small, task-focused cues composed of \textit{triggers} that initiate interface elements, \textit{rules} that govern responses, \textit{feedback} that communicates system state, \textit{loops} that govern recurring or time-based behavior, and \textit{modes} that adapt to behavior across situations \cite{saffer_microinteractions_2013}.
We apply this framework to SUIs, where AR-HMD cues must respond to evidence-based clinical needs without conflicting with patient care or team communication.
We introduced microinteractions and possible AR-HMD interaction modalities during co-design (Figure~\ref{fig:modalities}) to help HCWs reason about situated behavior; for example, task-specific timers may help clinicians anticipate time-sensitive interventions but become distracting when no longer relevant.
Prior healthcare design has also used iterative methods to refine interface microinteractions \cite{sonney_refinement_2022}.
Together, this framing supports using microinteractions in desktop-mediated spatial co-design to specify AR-HMD behavior.


\section{Methods}
\label{sec:methods}
\begin{figure*}[t]
\centering
\includegraphics[width=\linewidth]{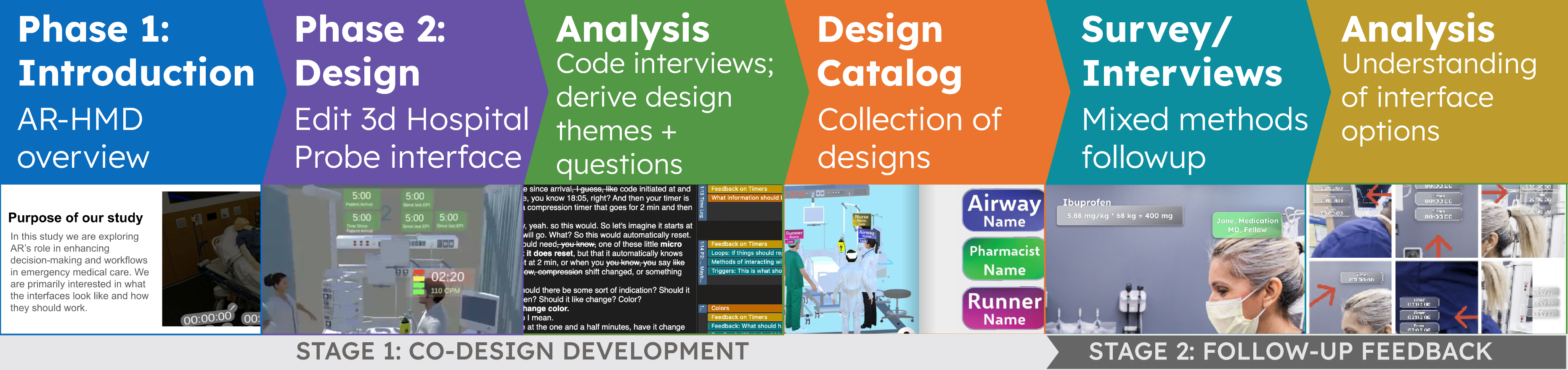}
\caption{Study flow: HCWs discussed ED coordination, refined AR-HMD concepts in an editable desktop-mediated Unity-based 3D design probe, and provided follow-up feedback on derived visual alternatives.}
\Description{Study flow with two stages. Stage 1 includes AR-HMD orientation, desktop-mediated co-design in an editable Unity-based 3D probe, transcript and artifact analysis, then synthesis of a visual design catalog. Stage 2 includes follow-up interviews plus a survey reviewing derived visual alternatives, followed by separate analysis of each feedback format.}
\label{fig:phases}
\end{figure*}

We conducted an Institutional Review Board (IRB)-approved formative study IRB0145631 examining how HCWs used remote, desktop-mediated 3D co-design to critique and refine WAI AR-HMD interfaces for ED teamwork.
The study had two stages (Figure~\ref{fig:phases}).
In Stage 1, 12 HCWs remotely co-designed AR-HMD interfaces for role awareness, task-specific timers, and dosage verification using an editable Unity-based 3D probe shared through desktop screen-sharing.
Following analysis, we synthesized Stage 1 concepts into a visual catalog of the resulting speculative design space.
Participants compared spatial interface behaviors against their accounts of existing practices, tools, and safety concerns.
In Stage 2, five returning participants completed follow-up interviews and 26 additional HCWs completed a survey reviewing visual alternatives derived from the initial co-design sessions.
In both stages, we treated the interfaces as speculative probes and analyzed feedback as formative design knowledge rather than evidence of clinical effectiveness, usability, or team performance.

\begin{table}[t]
\centering
\footnotesize
\caption{Participant Demographics}
\begin{tabular}{p{0.3cm} p{0.3cm} l p{0.3cm} p{0.3cm} p{2.5cm}}\hline
\textbf{ID} & \textbf{Age} & \textbf{Credentials} & \textbf{Exp.} & \textbf{AR} & \textbf{Hospital-Type} \\
\hline
P1 & 58 & Attending Physician, MD & 27 & 3 & NT \\
P2 & 45 & Nurse Practitioner, NP & 7 & 1 & T, NT, NP, FP \\
P3 & 39 & Attending Physician, MD & 9 & 3 & T, NT, U, NP \\
P4 & 37 & Attending Physician, MD & 5 & 3 & T, NT, U, NP \\
P5 & 38 & Attending Physician, MD & 5 & 3 & T, NT \\
P6 & 54 & Attending Physician, MD & 23 & 3 & T, NT, U \\
P7 & 43 & Attending Physician, MD & 9 & 2 & T, NT, U, NP, FP \\
P8 & 58 & Attending Physician, MD & 25 & 4 & T, NT, U, NP, FP \\
P9 & 43 & Registered Nurse, RN & 18 & 1 & T, U, NP \\
P10 & 29 & Resident Physician, MD & 3 & 4 & T, NT, NP \\
P11 & 31 & Attending Physician, MD & $>$1 & 3 & T, U, NP \\
P12 & 26 & Registered Nurse, RN & 3.5 & 2 & T, U, NP \\
\hline
\end{tabular}

{\small\raggedright
Exp: Experience in years; AR: Familiarity with AR-HMDs; hospital types:
T = Teaching, NT = Non-Teaching, U = Urban, NP = Non-Profit, and
FP = For-Profit. AR familiarity: 1 = Not familiar, 5 = Very familiar.\par}
\label{table:participants}
\end{table}

\subsection{Stage 1: Desktop-Mediated 3D Co-Design of Future AR-HMD Interfaces}
\label{sec:stage1}

In Stage 1, we used speculative co-design with microinteraction prompts to examine how HCWs critiqued and refined WAI AR-HMD behavior before development, allowing spatial concepts to be considered in relation to clinical practice and safety \cite{saffer_microinteractions_2013}.
Building on prior research identifying opportunities for AR-HMD assistance in ED teamwork \cite{taylor_co-designing_2025}, we focused on coordination challenges involving identifying responsibility for evolving care tasks, maintaining awareness of time-sensitive and recurring interventions, and accurately verifying medication dosages during rapid care.
We structured sessions around three \textbf{design goals (DG)}:
\begin{enumerate}
\item[\textbf{DG1}:] Role-based notifications that display clinicians' responsibilities as teams assemble to promote awareness of who is responsible for ongoing care tasks.
\item[\textbf{DG2}:] Task-specific timers that make time-sensitive interventions visible for coordination of upcoming or recurring actions.
\item[\textbf{DG3}:] Dosage verification interfaces that present dosage information for checking medication decisions during care.
\end{enumerate}
Participants could also propose concepts beyond these goals; thus, the goals served as starting points for how spatial information might support ED teamwork rather than fixed interface solutions.

\subsubsection{Participants.} We recruited 12 ED HCWs through emergency medicine listservs and referral-based outreach for one-hour co-design sessions compensated with an \$18 Amazon gift card. Participants included eight attending physicians, one resident physician, one nurse practitioner, and two registered nurses.
Participant ages ranged from 26 to 58 years ($M = 41.75$, $SD = 10.26$), and reported clinical experience ranged from more than one year to 27 years ($M = 11.29$, $SD = 8.95$).
Participants self-reported gender as women ($N = 5$) and men ($N = 7$).
Participants had worked across multiple hospital settings, including teaching (11), non-teaching (9), and urban (8) hospitals.
Participants reported low to moderate familiarity with AR-HMDs ($M = 2.4$, $SD = 1.02$) on a Likert scale from 1 (\textit{little to no experience}) to 5 (\textit{expert experience}) (Table~\ref{table:participants}).

\subsubsection{Design Probe.} Using an editable desktop-mediated Unity 3D probe depicting an ICU-style care room \footnote{\url{https://www.turbosquid.com/3d-models/icu-scene-model-1533489}}, we situated, rather than reproduced, speculative resuscitation scenarios involving team assembly, timed interventions, and dosage verification.
Participants viewed the Unity scene remotely through desktop screen-sharing and did not wear an AR-HMD or complete a simulated treatment task. Instead, we asked whether they had a specific clinical situation or procedure in mind and, if not, asked them to envision the focal AR-HMD interfaces during a resuscitation code.
During remote interviews, the research team repositioned, resized, added, or removed interface elements in response to feedback rather than presenting fixed sketches or slides.
This process enabled participants to critique WAI AR-HMD interfaces in relation to clinicians, patients, equipment, existing practices, and safety concerns without reproducing headset field-of-view constraints, embodied movement, depth perception, peripheral clutter, or live-care demands.
The refined designs later informed the Stage 2 visual alternatives.

\subsubsection{Procedure.} We conducted Stage 1 as one-hour, two-phase remote co-design interviews.
In \textbf{Phase 1}, we obtained consent and collected demographic responses (Table~\ref{table:participants}).
Next, we introduced the study goals and explained that participants could discuss AR-HMD interfaces for ED teamwork and guide refinements to spatial elements.
To establish a shared baseline across varying AR-HMD familiarity, we showed a HoloLens video excerpt (2:55--3:48).\footnote{\url{https://www.youtube.com/watch?v=po76IvgYHms&t=175s}}
Before Phase 2, we introduced the Unity environment as an ICU-style room with clinicians around a patient bed and showed examples of role-based notifications, task timers, and dosage verification.

In \textbf{Phase 2}, before the design probe critique, we introduced the microinteraction framework to discuss how spatial interface behavior could respond during care (Figure~\ref{fig:modalities}) \cite{saffer_microinteractions_2013}.
We used it as a prompt rather than requiring participants to adopt its terminology.
Next, participants identified a relevant situation, or envisioned a resuscitation code if none came to mind, and used it to critique and refine the Unity design probe.
For each focal interface area, we asked about experienced challenges, whether AR-HMDs could support the activity, which notifications would be useful and why, when they should appear, how clinicians should control or dismiss them, and what would make the interface useful during care (see Supplemental Materials).
The interviewer adjusted interface elements in response to feedback, translating participants' ideas into provisional AR-HMD elements for further critique.
This iterative cycle of situating a concept, specifying microinteraction behavior, revising the spatial artifact, and re-critiquing it informed the SCF-HMD framework presented in the Discussion.
Finally, participants reflected on conditions for safely evaluating the designs in future functional AR-HMD systems, including appropriate wearers, relationships to existing systems, and privacy or reliability concerns.

\subsubsection{Data Collection and Analysis.} We recorded audio and screen-sharing video from Stage 1 co-design sessions with participant consent.
We removed identifying information from transcripts and checked Zoom-generated transcripts against recordings to correct errors and link comments to Unity interface concepts revised during each session.
Two researchers conducted codebook thematic analysis \cite{braun_thematic_2012} of the Stage 1 transcripts.
Together, they developed a codebook spanning seven categories: motivation for AR, interface design specifications, medical situations, AR concerns, application-specific design, AR interaction, and participants' clinical roles.
Within AR interaction, we coded proposed behaviors using the microinteraction categories of \textit{triggers}, \textit{rules}, \textit{feedback}, \textit{loops}, and \textit{modes} introduced during the sessions \cite{saffer_microinteractions_2013}.
The researchers applied the codebook to transcripts, allowed utterances to receive multiple codes, and discussed discrepancies to refine theme definitions and agree on representative excerpts.
To assess inter-rater reliability for multi-label coding, we calculated Krippendorff's $\alpha$, yielding $\alpha = .79$.
Because interface elements changed during sessions, we interpreted participants' explanations alongside the corresponding Unity revisions.

For each interface concept, we coded \textit{triggers} when participants described when or how an element should appear or become actionable; \textit{rules} when they specified its behavior after activation; \textit{feedback} when they described how it should communicate; \textit{loops} when they described repeated behavior; and \textit{modes} when they described how behavior could adapt across ED teams and clinical situations.
We applied these codes across role-based notifications, task-specific timers, and dosage verification interfaces to identify recurring patterns and concept-specific concerns.
By comparing coded behaviors with corresponding Unity revisions and participants' explanations of clinical value or concern, we characterized formative specifications for WAI AR-HMD interfaces rather than evidence of clinical effectiveness or deployment safety.

\subsection{Stage 2: Follow-Up Feedback on Visual Interface Alternatives}
\label{sec:followup_study2}

After Stage 1, we synthesized visual alternatives from concepts proposed or refined during the Unity co-design sessions to examine whether HCWs converged on preferred configurations or identified context-dependent microinteraction requirements.
We collected formative feedback through interviews with returning participants and a survey of HCWs who had not participated in Stage 1, recruited through convenience sampling.
The alternatives addressed role-based notifications, task-specific timers, dosage verification, intervention tracking, and additional AR-HMD concepts.
We selected recurring or consequential Stage 1 variations in content, spatial placement, visibility, alerts, and verification behavior.

\subsubsection{Participants.} Stage 2 included 31 HCWs who provided uncompensated follow-up feedback: five Stage 1 participants returned for interviews, and 26 additional HCWs completed the survey.
Returning participants included two attending physicians, one nurse practitioner, and two registered nurses (Table~\ref{table:participants}).
Across both forms of feedback, participants included 17 attending physicians, six registered nurses, three nurse practitioners, two fellows, one resident physician, one physician assistant, and one child life specialist.
Participants were 25--34 years old ($N = 10$), 35--44 years old ($N = 13$), or 45--64 years old ($N = 8$).
Participants self-reported gender as women ($N = 16$), men ($N = 13$), non-binary ($N = 1$), or preferred not to disclose ($N = 1$).
Clinical experience included 0--5 years ($N = 7$), 5--10 years ($N = 10$), 10--15 years ($N = 4$), 15--20 years ($N = 5$), and more than 20 years ($N = 4$); one participant preferred not to disclose.
Participants reported experience across teaching, non-teaching, urban, non-profit, for-profit, and outpatient settings.

\subsubsection{Follow-Up Materials and Procedure.}
Returning participants reviewed visual alternatives derived from Stage 1, including a composite scenario depicting multiple AR-HMD elements during emergency care (Figure~\ref{fig:teaser}), and discussed SUI preferences, concerns, and fit with existing ED tools and coordination practices (see Supplemental Materials).
Survey participants reviewed corresponding alternatives in Qualtrics and ranked each set by dragging their most preferred option to the top and least preferred to the bottom.
The instrument examined configurations for role-based notifications, task-specific timers, dosage verification, and intervention tracking, and asked participants to rank additional AR-HMD concepts by perceived benefit.
Because some participants omitted ranking items, response counts varied across questions.
Together, the study examined whether Stage 1 alternatives remained appropriate across HCW perspectives and where their behavior might require contextual adaptation during care.

\begin{table}[t]
\centering
\footnotesize
\caption{Stage 2 response counts for ranking items; counts vary because
some participants omitted individual items.}
\label{tab:stage2_response_rates}
\begin{tabular}{@{}p{0.27\linewidth} c p{0.50\linewidth}@{}}
\toprule
\textbf{Ranking item} &
\textbf{Responses} &
\textbf{Alternatives} \\
\midrule
\raggedright Role information &
26 &
\raggedright Name, role, title. \tabularnewline
\raggedright Timer alerts &
19 &
\raggedright Visual and auditory feedback. \tabularnewline
\raggedright Dosage modality &
10 &
\raggedright Visual or auditory presentation. \tabularnewline
\raggedright Dosage content &
22 &
\raggedright Dose, equation, weight, history, timing. \tabularnewline
\raggedright Equation visibility &
18 &
\raggedright Shown once, retained, or omitted. \tabularnewline
\bottomrule
\end{tabular}
\end{table}

\begin{figure*}[t]
\centering
\includegraphics[width=\textwidth]{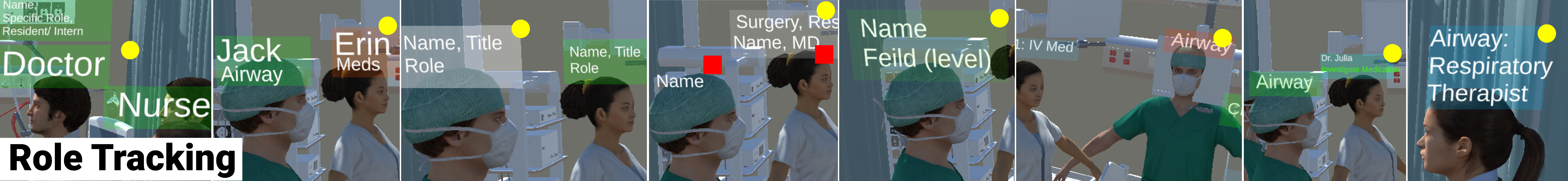}
\caption{Role-based notification alternatives refined through the
desktop-mediated design probe. Yellow circles indicate variations in
label content, clinician-linked placement, or role-based color coding;
red squares indicate minimized notifications.}
\Description{Unity-based role-based notification interface alternatives
placed above clinicians in a simulated emergency care environment.
Yellow circles mark alternatives involving different label content,
notifications positioned above clinicians, and color coding for roles or
responsibilities. Red squares mark reduced or minimized label designs.}
\label{fig:role_tracking}
\end{figure*}

\subsubsection{Data Collection and Analysis.} We recorded and transcribed follow-up interviews with participant consent.
We analyzed interviews and survey responses separately because the formats elicited different feedback, then compared them to identify where interview explanations supported, conflicted with, or qualified survey preferences.
We examined interview data for interface preferences, concerns, and fit with team dynamics and clinical workflows.
We also used Stage 1 visual designs as artifacts in an RtD-informed synthesis, treating them as objects for reflection and design knowledge production \cite{zimmerman_research_2007}.
Specifically, we reviewed the visual alternatives alongside interview explanations to organize recurring spatial placements, contextual variations, visibility modes, feedback mechanisms, and persistence behaviors into the design catalog.
In contrast, we analyzed survey and rank-order responses using first-choice frequencies and percentages, mean rank, and standard deviation among respondents to each item, and reviewed optional open-ended responses to contextualize  preferences.
We treated these data as formative feedback rather than evidence of usability, clinical effectiveness, or consensus on a final AR-HMD design.
We report descriptive statistics to characterize this formative sample, not for statistical inference or population-level generalization.


\section{Findings}
We organize the findings around three focal AR-HMD interfaces examined through a desktop-mediated 3D probe: role-based notifications, task-specific timers, and dosage verification, followed by broader concerns about AR-HMD use in ED environments.
Some Stage 2 surveys omitted ranking items, and the formative sample does not establish consensus, usability, or clinical effectiveness; instead, these findings refine the WAI design space within WAD constraints.
Participants' selected clinical situations shaped the qualities they emphasized.
For team assembly, participants focused on role notifications, name visibility, and whether labels should persist or fade.
For resuscitation codes, CPR, medication timing, or recurring interventions, they emphasized timers, urgency feedback, and minimizing interruptions to verbal coordination.
For medication decisions, they emphasized dosage verification, medication history, patient weight, and calculation transparency.
Thus, the alternatives represent context-dependent specifications that vary with clinical activity, team composition, and information available through existing tools, rather than universal preferences.

\subsection{Role-Based Notifications: Enhancing Visibility of Situational Awareness of Evolving Teams}

All participants envisioned role-based notifications as supporting awareness of team responsibilities during resuscitation.
Figure~\ref{fig:role_tracking} shows clinician-anchored notifications ranging from concise name-and-role labels to alternatives showing specialty, title, or current task.
P11 explained that a notification could show \textit{``Name, title so [\ldots], John, RN [\ldots] and then the role is [\ldots] [the medical task] you [are] assigned to, that can be on its own line.''}
P1 proposed notifications for clinicians with different medication-related responsibilities: \textit{``Some nurses might be handing out the medication and [other] nurses might be administering the medication. [\ldots] It'd be kind of helpful to know [\ldots] who's doing what [task] and where they're located.''}
Participants viewed these notifications as potential badge replacements for identifying unfamiliar team members and responsibilities, while still relying on verbal role assignment.

\begin{figure*}[t]
\centering
\includegraphics[width=\textwidth]{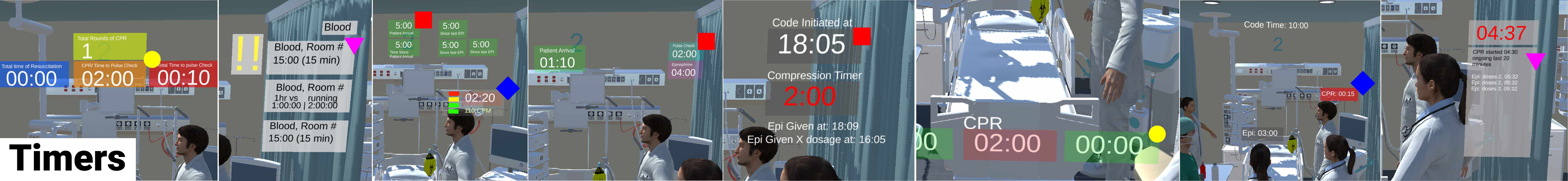}
\caption{Task-specific timer designs refined through the
desktop-mediated 3D design probe. Yellow circles indicate grouped
CPR-related timers; red squares indicate edge-of-view timer stacks;
blue diamonds indicate clinician-linked timers; magenta triangles
indicate room-event or medication-related timing contexts.}
\Description{Unity-based task-specific timer interface alternatives in
a simulated emergency care environment. Yellow circles mark grouped
CPR-related timer notifications. Red squares mark timers stacked along
the edge of the wearer's view. Blue diamonds mark timers positioned
above or near a relevant clinician. Magenta triangles mark timers for
other contexts, including room events and medication tracking.}
\label{fig:timers}
\end{figure*}

Participants specified \textit{triggers}, spatial \textit{rules}, and visibility \textit{modes} for changing teams and responsibilities.
P3 described an arrival-based \textit{trigger} and dismissible \textit{mode} as: \textit{``When the person walks in and scans in with their badge [\ldots] it auto populates their name and field [\ldots] the situation where you'd want to dismiss them is when you already know everyone on the team.''}
P6 specified a person-linked spatial \textit{rule} for cardiac-arrest roles, including airway, circulation, compressions, defibrillation, medication preparation, documentation, and timekeeping: \textit{``I've assigned him that role and then this will now just automatically populate above his head.''}
These person-linked notifications addressed a challenge in dynamically assembling teams: clinicians may need to identify unfamiliar colleagues and their tasks quickly, as P6 explained: \textit{``Teams often assemble from people who aren't necessarily people that I either work with immediately or even regularly [\ldots] they may include people who just responded to a call.''}

Five of 12 participants preferred notifications to remain visible throughout a code, while seven favored showing them during team assembly then fading them until a clinician arrived, a role changed, or the wearer requested them.
Five participants also proposed peripheral lists showing who had been called, was expected, or had temporarily left.
These lists formed a situational-awareness \textit{loop}, updating as team members joined, left, or changed tasks.
P6 explained: \textit{``I don't want to lose awareness of who my team members are and what they are potentially doing outside of the scope of what I see in front of me.''}
However, P6 warned about UI clutter in resuscitation rooms: \textit{``these [UI] overlays, [\dots][act as] callouts with people's head, are gonna get really cluttered.''}

Stage 2 feedback reinforced the need for concise, configurable notifications.
Name was ranked first by 22/26 respondents (84.6\%; $M = 1.42$, $SD = 1.06$), while role and title were each ranked first by 2/26 (7.7\%).
Interview participants differed on whether notifications should persist or appear only when needed.
Together, these findings suggest foregrounding identity while adapting visibility to team familiarity, task assignments, and visual clutter.

\subsection{Task-Specific Timers: Stateful Cues for Iterative, Time-Sensitive Tasks}

All participants discussed task-specific timers for CPR, medication administration, patient arrival, and other recurring interventions.
Figure~\ref{fig:timers} shows alternatives varying in placement, grouping, task association, and visual signaling.
Six participants preferred timers at the field-of-view edge or an anchored location rather than over the patient, as P2 described: \textit{``Or something that's stationary. I mean, they have a clock obviously on the wall, but something that is always constantly [\ldots][displayed] to the top right of their side.''}
P7 similarly envisioned grouped timers placed away from the center of care: \textit{``I would envision them all the time intervals grouped together, and either I kind of like it at the foot of the bed, where like there's not really anything else there.''}
Participants compared these alternatives with clocks, phone timers, and verbal timekeeping, suggesting AR-HMD timers could make task-specific timing glanceable without drawing attention away from care.

Participants specified timer \textit{triggers} indicating when elapsed time became relevant.
Patient arrival or code initiation could start broader care timers, while CPR or medication administration could trigger timers for pulse checks, compressor switches, repeat doses, or defibrillation.
Participants also proposed explicit requests to avoid unnecessary timers, displaying one only when relevant.
P11 proposed that a clinician could request a timer for a named event: \textit{``You can name the timer one of several things, pulse check, rhythm check, epinephrine, defibrillation [\ldots] and so the timer will only come up if you ask for it, and then, once it's up there, then you can be the one to monitor it.''}

Participants also specified \textit{feedback} and \textit{loops} for recurring tasks.
Six described losing track of timers or elapsed time during care, and five favored stateful interactions indicating when timers were nearing completion or had finished.
For a CPR compression timer, P2 specified anticipatory visual \textit{feedback}: \textit{``At the one and a half minutes, have it change color to red [\ldots] that way you have an idea [\ldots] that the recorder can then also say in 30 seconds you're going to switch [roles].''}
P2 also specified a recurring \textit{loop}: \textit{``And then your timer is compression is just a compression timer that goes for 2 min and then reset.''}
P2 also tied this reset to the clinical action: \textit{``Maybe you would need [\ldots] one of these little micro interactions so that it does reset, but that it automatically knows that it's going to reset at 2 min, or when you [\ldots] compression [\ldots] shift changed, or something like that.''}
P7 proposed flashing timers at meaningful intervals: \textit{``If flash is an option I think that that could even be nice [\ldots] at every 5 min interval, even like flash just to like draw your attention to like the passage of time [\ldots] the hardest thing is [\ldots] it's very easy in any of these scenarios to get task fixated and then you lose sight of like the global goal.''}

These findings frame task-specific timers as complementary, glanceable cues whose feedback signals urgency without unnecessarily interrupting care.
In Stage 2, flashing colors were most frequently ranked first for active timer \textit{feedback} (9/19, 47.4\%; $M = 1.58$, $SD = 0.61$), while auditory \textit{feedback} was most frequently ranked first for completed intervals (10/19, 52.6\%; $M = 1.68$, $SD = 0.82$).
However, interview participants cautioned that auditory feedback could add noise and interfere with team communication.

\begin{figure*}[t]
\centering
\includegraphics[width=\textwidth]{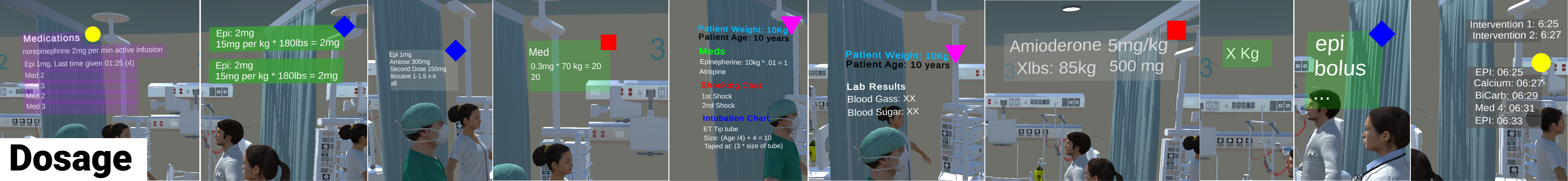}
\caption{Dosage verification interface designs refined through the
desktop-mediated design probe. Yellow circles indicate
medication-history displays; blue diamonds indicate shortened
medication lists; red squares indicate dosage-verification alternatives
exposing the calculation basis; magenta triangles indicate additional
patient or intervention context.}
\Description{Unity-based dosage verification interface alternatives in
a simulated emergency care environment. Yellow circles mark running
medication-history notifications. Blue diamonds mark shorter medication
lists. Red squares mark dosage verification alternatives that expose
the calculation basis. Magenta triangles mark notifications that include
additional patient or intervention information alongside medication
support.}
\label{fig:dosage}
\end{figure*}

\subsection{Dosage Verification Interfaces: Improving Reliable Decision Making}

Dosage verification prompted the clearest reframing of an initial concept: participants shifted from displaying a calculated dose toward helping clinicians verify it using medication history, patient weight, calculation visibility, and clinical context (Figure~\ref{fig:dosage}).
Four participants discussed hands-free verification and characterized medication history, patient weight, calculation visibility, and clinical indication as verification-focused microinteractions.

Patient weight and calculation visibility functioned as \textit{rules} for checking the output.
P3 explained: \textit{``Assuming that you already have the weight of the patient which I [\ldots][for an] adult would sometimes [be] a guessed weight there's not anything entered, but let's say you have the weight right? What I would say is [\ldots] ‘I think I would be okay with just the calculated dose man’. No, I would want the equation [visible]. So I would want, like 15 [milligrams] per kg times patients wait equals this.''}
P3 further described how the equation would support verification: \textit{``So that top line I could just look and say, okay, I want 2 milligrams then I could look below it and see the calculation there. That would allow me to verify that the weight actually listed is the weight that I want to consider.''}
Clinical indication provided contextual \textit{feedback} by clarifying why a dose was appropriate, as P5 explained: \textit{``epinephrine in a dose for cardiac arrest is different from [the dosage for] anaphylaxis,''} meaning the display needed to show which clinical situation the dose applied to.
Rather than replacing dosage references, medication histories, or clinician double-checking, participants valued AR-HMD support that made verification information easier to retrieve.

Participants envisioned dosage notifications connecting medication decisions to administration history without obstructing care or communication.
P11 proposed \textit{feedback} that, when \textit{triggered}, would appear near the administering clinician and then transition into a medication history: \textit{``Let's say you tell the nurse, ‘Please give one milligram of epinephrine.' I think having a little pop-up that says, ‘Give one milligram of epinephrine IV,' maybe near the nurse's hands, because that's when the medication is being given, would be helpful [\ldots] [and] disappear, and there could be a place where all of the medications that have been given show up. It'll say at what time, like, epinephrine was given.''}
P1 rejected auditory dosage \textit{feedback} because information should remain retrievable without competing with speech: \textit{``I wouldn't want somebody else talking and I miss something because this is talking.''}

Participants also identified appropriate moments for spatial dosage-verification information.
Five emphasized equations for rarely used medications, eight favored showing both the calculation and clinical indication, five envisioned populating information once relevant patient data were available, and six preferred medication lists minimized outside the direct field of view until needed.
For pediatric dosing, participants proposed an explicit verification \textit{loop} in which a virtual Broselow-style aid displayed estimated weight or color-zone information and required confirmation before accepting dosage information.
P2 explained: \textit{``Pediatric doses are weight based, and if you have a feature in there so that you automatically can see all of your doses. That would be extremely helpful. So you're almost like mimicking the PALS algorithm card.''}

Stage 2 feedback supported preserving calculation access while adapting administration timing, patient weight, and prior-dose information to clinical need.
Visual presentation was ranked first by 6/10 respondents (60.0\%; $M = 1.40$, $SD = 0.52$).
The equation received the most first-place rankings among dosage information options (6/22, 27.3\%; $M = 3.23$, $SD = 1.63$), while time since last administration had the strongest mean rank ($M = 2.45$, $SD = 1.18$).
Participants favored keeping the equation available for verification (13/18, 72.2\%; $M = 1.44$, $SD = 0.78$) over showing it once or omitting it.
Together, these findings frame dosage verification as clinician-controlled assistance that exposes calculation context, preserves medication history, and complements existing double-checking.
However, these concepts depend on accurately and rapidly retrieving patient weight, medication history, clinical indication, route, concentration, and administration time from institutional systems.
Thus, dosage verification is a formative requirement for transparent verification support, not evidence that AR-HMDs can provide reliable real-time medication guidance.

\subsection{Bridging Work-as-Imagined Interfaces and Work-as-Done Spatial Interface Constraints}
Beyond the three focal concepts, participants proposed interfaces for shared action logs, diagnostic checklists, medical records, vital signs, inventory guidance, EMS transitions, and documentation.
These concepts extended AR-HMD support from specific tasks toward broader access to patient status and team activity.
For example, P8 proposed supporting access to patient-monitor information during resuscitation: \textit{``It would be nice to cause that's like the main thing and then you would wanna see the monitor somewhere on this [\ldots] you want to visualize, like the patient's vital signs and things like that.''}
P5 similarly described a shared whiteboard where distributed team members could add information during a procedure: \textit{``Maybe there's a whiteboard there and then anyone who is part of this resuscitation is able to add bullet points. So for example, if child life [specialist] is outside and they manage to speak to the family, child life [specialist] can just go to the whiteboard.''}
Because these concepts received less discussion, we treat them as extensions of the WAI design rather than core findings; the design catalog provides additional details.\footnote{\url{https://heyzine.com/flip-book/5206d9f05b.htm}}

Participants also identified conditions shaping whether AR-HMD interfaces would be appropriate for ED teams.
A central constraint was who and how many HCWs should wear them, because wearer configuration determines what information is visible and whether it is available to the broader team.
All participants considered who should wear an AR-HMD, and five discussed multiple-wearer configurations (5/12).
P2 initially considered the recorder as a wearer, but then explained the value of broader visibility: \textit{``everyone should be able to [use it] [\ldots] and so then everybody kind of knows what they're doing.''}
Thus, coordination interfaces cannot assume universal headset use; they must communicate wearer-specific information to non-wearers or support multiple wearers while preserving shared awareness.

HCWs also raised privacy and reliability concerns that could limit when AR-HMDs should be used.
Three participants worried that patients might misunderstand the device or believe they were being recorded (3/12), as P4 explained: \textit{``Patients may be worried about their privacy [\ldots] are you recording me? What are you doing? Why are you wearing that thing?''}
Two participants raised concerns about dependence on AR-HMD assistance during critical activities (2/12); P6 cautioned: \textit{``this movement or me calling out something or me looking up at it needs to be really dependable.''}

\section{Discussion}
This study contributes SCF-HMD as a methodological framework for translating expert critique of WAI AR-HMD concepts into development goals for future spatial interfaces.
Using an editable desktop-mediated Unity 3D probe and microinteraction prompts, HCWs iteratively refined speculative concepts in relation to clinical practice and patient safety concerns.
The framework has three goals: 1) bridge WAI and WAD through design methods, 2) characterize spatial interface microinteractions for multimodal AR-HMD cues, and 3) specify what future prototypes should test through functional development.
Because participants viewed and discussed speculative interfaces rather than using a functional AR-HMD system, SCF-HMD should be understood as an early-stage design framework for eliciting testable interface requirements, not as evidence of usability, safety, clinical effectiveness, or improved team performance.

\begin{table*}[t]
\centering
\scriptsize
\caption{Speculative Co-Design Framework for AR-HMD Teamwork (SCF-HMD). The framework supports framing design problems, refining spatial concepts through microinteraction critique, and translating outputs into evaluation goals.}

\setlength{\tabcolsep}{4pt}
\renewcommand{\arraystretch}{1.05}

\begin{tabular}{
>{\raggedright\arraybackslash}p{2.8cm}
>{\raggedright\arraybackslash}p{4.1cm}
>{\raggedright\arraybackslash}p{5.0cm}
>{\raggedright\arraybackslash}p{4.4cm}
}
\toprule
\rowcolor{gray!25}
\textbf{Step} &
\textbf{Methods} &
\textbf{Research Action} &
\textbf{Contribution / Output} \\
\midrule

\rowcolor{gray!18}
\textbf{Phase 1:} & \textbf{Frame the problem} & & \\

\textbf{Identify team challenge}
& Interviews, focus groups, and observations
& Identify coordination problems, existing tools, and safety constraints
& Ground the problem in stakeholder experience \\

\textbf{Bound intervention}
& Participatory and speculative design, design fiction, and WAI/WAD
& Define how speculative AR-HMD interfaces complement current practice
& Identify where spatial interfaces may help or create risk \\

\textbf{Construct design probe}
& RtD, spatial probes, 3D mockups, and storyboards
& Build or adapt a 3D environment with concepts around people, tasks,
and equipment
& Create a situated 3D probe for critique \\

\midrule

\rowcolor{gray!18}
\textbf{Phase 2:} & \textbf{Iterate through co-design} & & \\

\textbf{Reflect on team scenarios}
& Scenario elicitation and expert walkthroughs
& Prompt discussion around a selected or provided care scenario
& Relate concepts to workflow and local constraints \\

\textbf{Specify spatial microinteractions}
& Microinteraction framework
& Examine how interfaces appear, update, or are dismissed
& Specify \textit{triggers}, \textit{rules}, \textit{feedback},
\textit{loops}, and \textit{modes} \\

\textbf{Revise and re-critique}
& Live prototyping, co-design, surveys, and storyboarding
& Revise placement, content, or behavior during interviews
& Refine concepts for communication, awareness, verification, and safety \\

\midrule

\rowcolor{gray!18}
\textbf{Phase 3:} & \textbf{Synthesize and prepare evaluation} & & \\

\textbf{Analyze revisions}
& Thematic and artifact analysis
& Analyze transcripts, artifact changes, and specifications
& Identify requirements, tensions, concerns, and limitations \\

\textbf{Synthesize artifacts}
& Design catalogs and storyboards
& Create visual alternatives, catalogs, or follow-up artifacts
& Preserve concepts for critique, extension, and comparison \\

\textbf{Develop prototype}
& Stakeholder review, user studies, simulation, and deployment
& Translate specifications into functional prototypes
& Define targets for system evaluation \\

\bottomrule
\end{tabular}
\Description{A four-column table describing the Speculative Co-Design Framework for AR-HMD Teamwork (SCF-HMD). The table is organized into three phases: frame the design problem, iterate through desktop-mediated co-design, and synthesize and prepare evaluation. Each phase includes steps, methods, research actions, and resulting contributions or outputs.}
\label{table:scf_hmd}
\end{table*}

\subsection{Speculative Co-Design Framework for AR-HMD Teamwork (SCF-HMD)}
Healthcare AR-HMD research has demonstrated opportunities for presenting clinical information to individual wearers, and research on time-sensitive teamwork emphasizes that new interfaces must avoid adding distraction or coordination burden \cite{kimmel_opticare_2021, tanjim_help_2025, taylor_rapidly_2025}.
Building on this prior work, we contribute the \textbf{Speculative Co-Design Framework for AR-HMD Teamwork (SCF-HMD)}, a methodological design framework for bridging WAI and WAD before functional implementation.
SCF-HMD guides researchers in using editable desktop-mediated probes and microinteraction prompts to iteratively refine low-fidelity AR-HMD concepts, characterize multimodal spatial interface behavior, and specify development goals for later prototyping and evaluation. 
Rather than assuming that AR-HMDs improve teamwork, SCF-HMD begins with an existing coordination challenge and examines whether a speculative spatial intervention could complement current practice.

Table~\ref{table:scf_hmd} presents this process across three phases.
First, researchers can \textit{frame the SUI design problem} as identifying a team challenge, bounding the intervention in relation to existing tools and WAI/WAD, and constructing an editable desktop-mediated probe situated in a recognizable work environment.
Researchers can translate domain challenges into spatial designs placed around people, tasks, and equipment, to produce a situated design probe that can be critiqued before implementation.
Second, \textit{through iterative refinement of SUI co-design artifacts}, researchers can prompt experts to relate domain challenges to experiential scenarios, specify SUI microinteractions, and critique artifacts through live revisions. This process generates situated microinteraction specifications that reflect participant concerns, including, in our case, team communication, shared awareness, verification, and safety.
This includes examining how interfaces appear, update, provide feedback, form loops, or are minimized and dismissed.
Third, using SCF-HMD, researchers can \textit{synthesize and prepare evaluation protocols} through analysis of interview transcripts, artifact revisions, and microinteraction specifications to introduce a final interface, including design catalogs, visual alternatives, and development goals that identify future sensing, perception, and multimodal interaction cues in SUI.

In our study, for example, participants used our design probe to reconsider a calculated-dose display as a verification-oriented concept that exposed patient-specific information and remained available until dismissal.
This illustrates the benefit of SCF-HMD over other design approaches that only document stakeholder needs or collect preferences–that is, our framework helps translate expert critique into spatial and temporal interface behaviors that developers can later prototype and test.

SCF-HMD, therefore, contributes a methodological design framework rather than a fixed set of AR-HMD interface requirements.
Its outputs can inform, rather than replace, later implementation and evaluation protocols.
Furthermore, it facilitates early-stage development of SUIs through the translation of WAI concepts into testable development goals grounded in participants' accounts of WAD, which future simulation and deployment studies can compare against existing clinical practices and tools.

\subsection{Desktop-Mediated 3D Co-Design for Work-as-Imagined SUIs}
Our desktop-mediated 3D probe builds on participatory and speculative research using spatial representations to critique technologies before deployment \cite{mcveigh-schultz_immersive_2018, chi_participatory_2022, taylor_hospitals_2022}.
Unlike in-person or headset-based prototyping, it allowed HCWs to participate remotely while considering placement relative to clinicians, patients, equipment, and care.
It moved HCWs from broad AR-HMD ideas to situated judgments about interfaces' fit with or disruption of clinical work.

However, this method should not be interpreted as true spatial immersion. Participants viewed a screen-shared Unity scene on a desktop while the research team manually manipulated the 3D environment in response to verbal feedback. The probe therefore supported spatial reasoning about placement, proximity, and interface behavior, but did not reproduce headset field-of-view limits, embodied movement, true 3D depth perception, realistic scale, peripheral clutter, or live-care visual demands. Its contribution is formative: it supports refining spatial interface concepts before implementation and evaluation. The visual design catalog preserves these concepts and concerns for future SUI research.

\subsection{From Individual Critique to Team-Level Evaluation}
Although this study examines AR-HMD interfaces for ED teamwork, the co-design sessions and follow-up feedback were conducted with individual HCWs. This was appropriate for eliciting early-stage critique of speculative interface concepts, but it cannot capture the synchronous communication, shared situational awareness, and social dynamics of multiple HCWs coordinating in the same room. An interface that appears clean or useful to an individual participant viewing a desktop display may create different problems in live team settings, including visual clutter, blocked eye contact, interruptions to closed-loop communication, or friction between headset wearers and non-wearers.

These concerns are especially important for ED teamwork because spatial information is not only an individual display problem, but also a coordination problem. Role-based notifications, timers, and dosage verification interfaces may need to support shared reference points across team members rather than optimizing a single wearer's view. Future work should therefore evaluate selected concepts in multi-user simulations or team-based co-design sessions, where researchers can examine how AR-HMD cues affect verbal communication, gaze, handoffs, shared awareness, and the ability of teams to coordinate without adding new workload or obstruction.

\subsection{Institutional Infrastructure and Real-Time Data Constraints}
Participants discussed AR-HMD interfaces as if relevant information could appear when needed, but deployment would depend on institutional infrastructure beyond this study’s scope. Dosage verification, medication history, intervention tracking, patient weight, role assignment, and timing information would likely require integration with EHRs, medication administration records, automated dispensing systems, patient monitors, badge or location systems, and local documentation practices. These integrations introduce constraints involving data availability, latency, reliability, authentication, permissions, and provenance.

These constraints could change the interface requirements identified in this study. For example, a dosage verification display may be useful only if it clearly indicates the source, timestamp, and uncertainty of patient-specific information. A timer or intervention-tracking display may need to distinguish between automatically sensed events, manually entered events, and unverified information. During a crisis, delayed or incomplete backend data could be worse than no display if clinicians treat the interface as authoritative. Future AR-HMD prototypes should therefore test not only spatial placement and microinteraction behavior, but also how systems communicate data source, freshness, uncertainty, and clinician control within real hospital infrastructure.

\subsection{Limitations and Future Work}
This study has several limitations. First, participants critiqued a desktop-mediated Unity 3D probe rather than a functional AR-HMD system in clinical practice or simulation. Although the probe supported spatial reasoning about interface placement and behavior, it did not reproduce embodiment, depth perception, field-of-view constraints, peripheral clutter, sensing errors, workload, or the pressure of real patient care. Second, both stages elicited feedback from individual HCWs rather than live clinical teams, so the study cannot capture synchronous team dynamics, closed-loop communication, eye contact, or the friction that may occur when multiple HCWs coordinate around a patient.

Third, the sample was weighted toward physicians, including nine of 12 participants in Stage 1 and 20 of 31 in Stage 2. This limits representation of nurses, respiratory therapists, technicians, and other HCWs central to ED teamwork. A more interprofessionally balanced cohort might produce different requirements for notification persistence, medication administration support, documentation, equipment coordination, and tolerance for visual or auditory interruptions. Finally, follow-up interviews and survey rankings elicited different but complementary feedback, and some survey participants omitted ranking items. We therefore treat the findings as formative refinements of speculative concepts rather than consensus on final designs or evidence of usability, safety, or clinical effectiveness. Future work should apply SCF-HMD with broader interprofessional teams and evaluate concepts in simulation.

\section{Conclusion}

Through speculative co-design with HCWs, we examined future AR-HMD interfaces for safety-critical ED teamwork.
Using a desktop-mediated Unity 3D probe and microinteraction prompts, participants translated role-based notifications, task-specific timers, and dosage verification concepts into specifications grounded in clinical practice and safety concerns.
We proposed the SCF-HMD, a reusable process for eliciting and refining work-as-imagined spatial interface behavior before functional implementation. 
Rather than demonstrating the usability or effectiveness of a deployed AR-HMD system, SCF-HMD supports SUI and HCI research in developing AR-HMD systems that complement existing high-stakes teamwork rather than assuming augmentation is inherently beneficial.

\begin{acks}
This work was supported in part by National Science Foundation (NSF) Awards 2439474, 2444132, 2327569, 2238313, and 2223432.
\end{acks}

\bibliographystyle{ACM-Reference-Format}
\bibliography{references}

\end{document}